\documentclass{article}
\usepackage[T1]{fontenc} % add special characters (e.g., umlaute)
\usepackage[utf8]{inputenc} % set utf-8 as default input encoding
\usepackage{ismir,amsmath,cite,url}
\usepackage{graphicx}

\usepackage{color}

\usepackage[firstpage]{draftwatermark}
\definecolor{lightgray}{rgb}{0.9,0.9,0.9}
\definecolor{darkgray}{rgb}{0.4,0.4,0.4}
\SetWatermarkFontSize{12pt}
\SetWatermarkScale{1.1}
\SetWatermarkAngle{90}
\SetWatermarkHorCenter{202mm}
\SetWatermarkVerCenter{170mm}
\SetWatermarkColor{darkgray}
\SetWatermarkText{Late-Breaking / Demo Session Extended Abstract, ISMIR 2024 Conference}

\usepackage{makecell}
\usepackage{multirow}
\def\lbd{}	% Flag to use correct LBD settings in the paper, please do not modify this line

\title{BeatlesFC: Harmonic function annotations of Isophonics' The Beatles dataset}

\threeauthors
  {Ji Yeoung Sim} {CUNY Graduate Center \\ Music \\ { }}
  {Rebecca Moranis} {CUNY Graduate Center \\ Music \\ { }}
  {Johanna Devaney} {Brooklyn College and CUNY Graduate Center  \\ Data Analysis \& Visualization and Music \\ {\tt johanna.devaney@brooklyn.cuny.edu}} 

\def\authorname{J.Y. Sim, R. Moranis, and J. Devaney}

\begin{document}

\maketitle
\begin{abstract}
This paper presents BeatlesFC, a set of harmonic function annotations for Isophonics' The Beatles dataset. Harmonic function annotations characterize chord labels as stable (tonic) or unstable (predominant, dominant). They operate at the level of musical phrases, serving as a link between chord labels and higher-level formal structures.

\end{abstract}
\section{Introduction}\label{sec:introduction}

The Isophonics' reference annotations\footnote{http://www.isophonics.org/datasets} feature a multilayered labeling system \cite{harte2010towards} that includes both lower-level analyses, such as individual chords and beat types, and higher-level analyses, such as key and song structure. We present BeatlesFC, a set of harmonic function annotations for  Isophonics' The Beatles dataset, to provide a link between these levels. Although the harmonic function analysis has been included in symbolic music datasets of Western art music (e.g., \cite{devaney2015theme,tymoczko2019romantext, gotham2023rome}), it has yet to be widely included in audio-based popular music datasets.

%However, it does not include higher-level labels that convey the function of individual harmonies, which cannot give much information about musical context on a larger scale for popular music. I

%\cite{de2011corpus,harte2010towards,burgoyne2011expert}.

%As harmonic function labels reflect not only the unique harmonic syntax of rock but also the various functions of each chord at the phrase level, they can provide additional information about the musical context on a larger scale. As a demonstration model, the harmonic function annotations were based on 179 songs in The Beatles dataset from the Isophonics, chosen for the dataset's compatibility with the analytical framework and the quality of the existing labels.

%making it actively used in research on Automatic Chord Recognition (ACR) related to popular music \cite{harte2010towards}. 

\section{Analytical Approach}

\subsection{Nobile's Function Circuit Approach}
The harmonic function annotations in Beatles FC are based on Nobile's concept of the functional circuit in rock music \cite{nobile2016harmonic}. The fundamental assumption of this analytic approach is that  harmonies in rock typically move from stable, tonic (T), to unstable, predominant (PD) or dominant (D), harmonies, and back to stable ones. It is rooted in Schenkerian analysis of Western art music, where the large-scale harmonic structure serves as the foundation from which smaller progressions emerge. However Nobile's function circuit approach is designed for rock music rather than Western art music. For example, while in Western art music it is common for the dominant function to be tied to the dominant chord (V), the dominant function in the function circuit approach allows for various chords to fill a dominant function, reflecting the unique harmonic practices in rock music. %Therefore, functional circuit analysis lies in its emphasis on establishing the harmonic syntax of rock itself. %By employing the functional circuit as our main theoretical framework for harmonic function analysis, we assumed that we could generate more proper harmonic function labels that focus on original harmonic progression in rock practice.  
Function circuit analyses demonstrate how sequence of chords operates together as a larger single harmonic unit, specifically through the use of prolongation techniques like neighboring chords and back-related chords.

\subsection{The Beatles Dataset}
The Beatles' songs for are particularly well suited for function circuit approach because they are structured either in the strophic or AABA form with clear cadences, which is characteristic of the rock music of the early- and mid-1960s.
%which showcases the distinctive harmonic progression between the thematic A section and the contrasting B section, and clear cadences. % Also, The Beatles's music holds significant influence within the pop and rock genres, as demonstrated by their album \textit{Sgt. Pepper’s Lonely Hearts Club Band} being selected for the \textit{Cambridge Music Handbook}series, which primarily features classical masterpieces. \cite{spicer2001british}.

\begin{figure*}[ht]
    \centering
    \includegraphics[width=.9\textwidth]{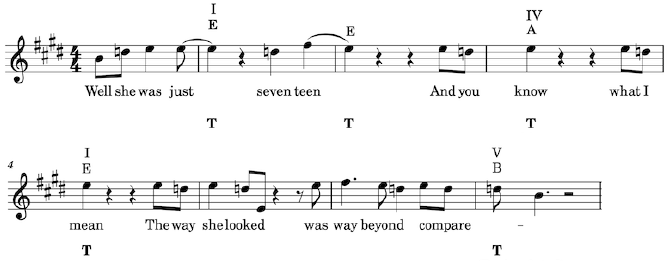}
    \caption{An analysis example from an excerpt of ``I Saw Her Standing There'' by The Beatles. The pop chord symbols are the original Isophonics chord annotations, the Roman numerals are calculated from the Isophonics chord and key annotations, and the harmonic function labels are the new contribution of this dataset.}
    \label{fig:analysis}
\end{figure*}

\section{Annotation Process}

Two PhD-level music theory students created the annotations. Each annotator has several years of experience teaching harmonic analysis and has taken one or more theory courses on popular music. This background ensures a shared understanding of functional circuit analysis. The annotators divided the songs between them (Annotator A analyzed 126 songs and Annotator B analyzed 79 songs), and double-checked the quality of annotations by inter-annotating two songs in each album (26 of the above counts were cross-checked). Sonic Visualizer was used to listen to the audio in relation to the original Isophonics chord label annotations. %In Sonic Visualizer, the annotators opened the original chord label file from the Isophonics dataset and changed 
A function label (T, PD or D) was assigned for each chord label in the original annotations in order to maintain the same onset and offset as the chord labels.
% to be compatible with the structure of the existing Isophonics dataset. 
In addition to chord labels from the original dataset, the annotators considered the Isophonics key and structural segment labels to interpret the chord labels within the broader harmonic context of each song.

\section{Function Annotations}

In total, 179 harmonic function annotations were created out of the 180 total songs in Isophonics' A discernible key is necessary as the pop chord labels are interpreted with respect to key in order to assess their harmonic function. In cases of modulation, the annotators considered the Isophonics key labels as a local key and analyzed the function label in the key area. The Beatles dataset. One song ``Revolution 9'' from \emph{The Beatles} album was excluded, as it lacks any discernible key. We, however, opted to annotate the three raga-influenced songs, ``Love to You'', ``I Want to Tell You'', and ``Within You Without You'', despite the likely issues with applying the function circuit theory these songs, in order to examine where the utility of this analytic method can break down. 

% “07-04-Love You To”, “07-12-I Want to Tell You”, “08-08-Within_You_Without_You” are the raga-influenced songs

Occasionally, the annotators disagreed on the functions. We have preserved the separate annotations for researchers interested in examining persistent points of disagreement (an example of this is discussed in Section \ref{example} below).  Each completed function-label analysis was saved as a .lab file with the start and stop times (in seconds) of each harmonic function. The complete dataset is available for download from GitHub\footnote{\url{https://github.com/jcdevaney/beatlesFC}}.

%In cases of disagreement, the annotators discussed their interpretations and reached a consensus, where possible.

\subsection{Statistics}
Out of the 14,132 labels in the 179 songs, the T function appears most frequently, with 9,941 occurrences, accounting for approximately 70.3\% of the total number of labels. The PD function is the second most common, with 2,326 occurrences (16.5\%), and the D function is the least common, with 1,865 occurrences (13.2\%). The underlying assumption in the function circuit analysis is that rock harmonies transition from stable chords (T) to unstable ones (PD and D) and then return to more stable chords (T). The T labels occurring at roughly twice the frequency of the combined predominant and dominant function labels demonstrate that the The Beatles’ songs reflect rock’s harmonic syntax of the functional circuit.

\subsection{Analytical Example}
\label{example}
Figure \ref{fig:analysis} shows an analytical excerpt from ``I Saw Her Standing There'' from The Beatles' first album \emph{Please Please Me}. In ``I Saw Her Standing There', the 16-bar blues progression consists of the verse, which is built upon the three small subsections: statement, departure, and cadence. The excerpt exhibits one of the verse's subsections called a statement. The statement of the 16-bar blues progression can be interpreted as maintaining the tonic \cite{nobile2020form}. The first chord E shows a T function and it lasts until measure 8, which is the end of the statement. Throughout the 8 measures, chords maintain a T function using prolongation techniques that reflect rock harmonic practice. In measure 3, the A chord prolongs the T function as a neighboring chord between two E tonic chords instead of establishing a new subdominant function. The B chord in measure 8 also can be seen as a tonic prolongation chord with a back-related V rather than displaying a dominant function. This excerpt also serves as example of inter-annotator disagreement as the second annotator saw the V in measure 8 as a tonic prolongation with the back-related V chord. The differing interpretations have been preserved in the '-A' and '-B' versions of the annotations in the repository. 

%The consideration of the harmonic function at a phrase level assures the accuracy and coherence of this annotation. 

% \begin{figure}
%     \centering
%     \includegraphics[width=1.0\linewidth]{graph1.png}
%     \caption{the }
%     \label{fig:enter-label}
% \end{figure}

% \begin{table}[]
%     \centering
%     \begin{tabular}{c|ccc}
%          Total & Tonic (T) & Predominant (PD) & Dominant (D)\\
%          \hline
%          14132 & 9941 & 2326 & 1865 
%     \end{tabular}
%     \caption{Tally of the number of labels with each function.}
%     \label{tab:my_label}
% \end{table}

\section{Conclusion}
This paper has introduced BeatlesFC, which we believe to be the first harmonic function annotations for popular music. Harmonic function annotations provide a link between lower-level annotations, such as chord, and higher-level annotations, such as key and song structure. We annotated harmonic function in Isophonics' The Beatles dataset because it contains annotations at both the lower- and higher-levels and because the music of The Beatles is particularly well suited for the function circuit analytic approach we employed.

\section{Acknowledgments}
This work is supported by the National Science Foundation (NSF) grant 2228910. The opinions expressed in this work are solely those of the authors and do not necessarily reflect the views of the NSF.

%Second, harmonic function annotations capturing the harmonic structure of the AABA form can provide musical context for the Automatic Chord Recognition system at the phrase level. Finally, the annotations will be useful for the observation of changes in the artist's harmonic syntax across their output. Employing our harmonic function annotations for ACE is our plan for future work.

%\newpage
% For bibtex users:
\bibliography{ISMIRtemplate}

% For non bibtex users:
%\begin{thebibliography}{citations}
% \bibitem{Author:17}
% E.~Author and B.~Authour, ``The title of the conference paper,'' in {\em Proc.
% of the Int. Society for Music Information Retrieval Conf.}, (Suzhou, China),
% pp.~111--117, 2017.
%
% \bibitem{Someone:10}
% A.~Someone, B.~Someone, and C.~Someone, ``The title of the journal paper,''
%  {\em Journal of New Music Research}, vol.~A, pp.~111--222, September 2010.
%
% \bibitem{Person:20}
% O.~Person, {\em Title of the Book}.
% \newblock Montr\'{e}al, Canada: McGill-Queen's University Press, 2021.
%
% \bibitem{Person:09}
% F.~Person and S.~Person, ``Title of a chapter this book,'' in {\em A Book
% Containing Delightful Chapters} (A.~G. Editor, ed.), pp.~58--102, Tokyo,
% Japan: The Publisher, 2009.
%
%
%\end{thebibliography}

\end{document}